\documentclass[11pt,a4paper,english,preprint,aps,eqsecnum,superscriptaddress]{revtex4}
\usepackage{amsmath}
\usepackage{graphicx}
\usepackage{amssymb}
\makeatletter
\DeclareMathOperator{\tr}{Tr}

\usepackage{babel} \makeatother
\newcommand{\dk}{\frac{d^{4}k}{\left(2\pi\right)^{4}}}
\newcommand{\dt}{d^{4}\theta}
\newcommand{\dth}{\mathbb{D}_{\theta}}

\begin{document}

\title{Towards a consistent noncommutative supersymmetric Yang-Mills theory: superfield covariant analysis.}

\author{A. F. Ferrari}
\author{H. O. Girotti}
\affiliation{Instituto de F\'{\i}sica, Universidade Federal do Rio Grande do
Sul, Caixa Postal 15051, 91501-970 - Porto Alegre, RS, Brazil}
\email{alysson, hgirotti, aribeiro@if.ufrgs.br}

\author{M. Gomes}
\author{A. Yu. Petrov}
\altaffiliation[]{ Department of Theoretical Physics,
Tomsk State Pedagogical University
Tomsk 634041, Russia
(email: petrov@tspu.edu.ru)}
\affiliation{Instituto de F\'{\i}sica, Universidade de S\~{a}o Paulo, Caixa
Postal 66318, 05315-970, S\~{a}o Paulo - SP, Brazil}
\email{mgomes, petrov, rivelles, ajsilva@fma.if.usp.br}

\author{A. A. Ribeiro}
\affiliation{Instituto de F\'{\i}sica, Universidade Federal do Rio Grande do
Sul, Caixa Postal 15051, 91501-970 - Porto Alegre, RS, Brazil}

\author{V. O. Rivelles}
\author{A. J. da Silva}
\affiliation{Instituto de F\'{\i}sica, Universidade de S\~{a}o Paulo, Caixa
Postal 66318, 05315-970, S\~{a}o Paulo - SP, Brazil}

\begin{abstract}
Commutative four dimensional supersymmetric Yang-Mills (SYM) is known to be renormalizable for
${\mathcal N} = 1, 2$, and finite for ${\mathcal N} = 4$. However, in the
noncommutative version of the model the UV/IR mechanism
gives rise to infrared divergences which may spoil the perturbative
expansion. In this work we pursue the study of the consistency of the ${\cal N} = 1, 2, 4$
noncommutative supersymmetric Yang-Mills theory with gauge group $U(N)$ (NCSYM). We employ the covariant superfield framework to compute the one-loop corrections to the two- and three-point functions of the gauge superfield $V$. It is found that the cancellation of the harmful UV/IR infrared divergences only takes place in the fundamental representation of the gauge group. We argue that this is in agreement with the low energy limit of the open superstring in the presence of an external magnetic field. As expected, the planar sector of the two-point function of the $V$ superfield exhibits UV divergences. They are found to cancel, in the Feynman gauge, for the maximally extended ${\cal N} = 4$ supersymmetric theory. This gives support to the belief that the ${\cal N} = 4$ NCSYM theory is UV finite.
\end{abstract}

\maketitle

\newpage 

\section{Introduction}
\label{sec:level1}

The four dimensional maximally extended supersymmetric Yang-Mills
theory has recently been under intense scrutiny particularly due to
its connection with the string/brane theory 
(see, for example, \cite{Kovacs} and references therein). By
contrast, studies on the renormalization properties of its
noncommutative counterpart are more scarce.  Some 
indications of its ultraviolet finiteness have already been given:
using the background field formalism the one-loop four point function
of the field strength has been computed \cite{Zanon1,Zanon2} and
the finiteness of the model has been argued on the basis of the
vanishing of the relevant beta function \cite{JJ}. However, a direct
calculation of the Green functions of the gauge potential $V$ within the
superfield formalism is still lacking. As known, this study is
essential to secure the ultraviolet finiteness of the higher 
order contributions to the effective action. Besides, the possible
existence of nonintegrable infrared singularities, which can jeopardize
the perturbative expansion, is particularly worrisome. Here we will furnish 
some further insight into these questions
by providing an explicit calculation of the one-loop corrections to the
two- and three-point vertex functions of the $V$ superfield.
This work is the follow up of a recent paper of us dealing with the
divergence structure of noncommutative supersymmetric
QED$_4$ \cite{Ferrari1}. We shall use again the covariant
superfield formalism to study the analogous problem for the
${\mathcal N } = 1, 2, 4$ NCSYM theories in four
space-time dimensions \cite{footnote1}. The reason for choosing $U(N)$ as the gauge group is based on the fact that it admits a simple noncommutative extension \cite{Chaichian,footnote3}. 

In Sec.  \ref{sec:level2} we present the action and the Feynman rules deriving from it, for the NCSYM theory. The one-loop corrections to the two-point function of the gauge superfield $V$ are computed and discussed in Sec.  \ref{sec:level3}. As will be seen, the fact that the gauge group is enforced to be in the fundamental representation warrants the cancellation of the leading UV/IR infrared divergences. We also demonstrate that, in the case ${\mathcal N} = 4$, the planar sector of the one-loop corrections to the two-point function of the $V$ superfield turn out to be free of UV divergences. This supports the belief that, as in the commutative case, the ${\mathcal N} = 4$ NCSYM theory is UV finite. 

Consistency demands that the cancellation of the nonintegrable UV/IR infrared divergences should also take place in the fundamental representation for any n-point vertex function. This is, indeed, verified, in Sec.  \ref{sec:level4}, for the leading UV/IR infrared divergences in the one-loop corrections to the three-point vertex function of the $V$ superfield. The conclusions and some final remarks are contained in Sec.  \ref{sec:level5}.

\section{The Action and Feynman rules of NCSYM}
\label{sec:level2}

In ${\mathcal{N}}=1$ superspace, NCSYM in four space-time dimensions is described by the
nonpolynomial action

\begin{equation}
S_{V}\,=\,-\,\frac{1}{2g^{2}}\int
d^{8}z\,\tr\,\left(e^{-gV}*D^{\alpha}e^{gV}\right)*\overline{D}^{2}\left(e^{-gV}*D_{\alpha}e^{gV}\right)\,,\label{I-1}
\end{equation}

\noindent 
where $g$ is the coupling constant, $V$ is a real Lie algebra
valued vector superfield,

\begin{equation}
V\left(z\right)\,=\, V_{a}\left(z\right)\,
T_{a}\,,\label{I-2}
\end{equation}

\noindent 
and the $T_{a}, a = 1, \cdots, N^2$, are the generators of $U(N)$ in the fundamental representation. They satisfy the algebra

\begin{equation}
\left[T_{a}\,,\, T_{b}\right]\,=\, i\, f_{abc}\, T_{c}\,.\label{I-3}
\end{equation}

\noindent 
Here, the $f_{abc}$'s are the structure constants of the gauge group. The generators
are normalized according to

\begin{equation}
\tr\left(T_{a}T_{b}\right)\,=\,\frac{1}{2}\,\delta_{ab}\,.\label{I-4}
\end{equation}

\noindent 
Furthermore, $D^2 = 1/2 D^{\alpha} D_{\alpha}$, ${\bar D}^2 = 1/2 {\bar D}_{\dot{\alpha}} {\bar D}^{\dot{\alpha}}$ and the Moyal product of field functions is defined as usual, i.e.,

\begin{eqnarray}
 &  & \phi_{1}(x)\ast\phi_{2}(x)\,=\,\phi_{1}(x)\,
exp\left(\frac{i}{2}\,\overleftarrow{\frac{\partial}{\partial
x^{\mu}}}\,\Theta^{\mu\nu}\,\overrightarrow{\frac{\partial}{\partial
x^{\nu}}}\right)\,\phi_{2}(x)
\nonumber \\ &  &
=\,\sum_{n=0}^{\infty}\,\left(\frac{i}{2}\right)^{n}\,\frac{1}{n!}\,\left[\partial_{\mu_{1}}\partial_{\mu_{2}}...\partial_{\mu_{n}}\,\phi_{1}(x)\right]\,\Theta^{\mu_{1}\nu_{1}}\Theta^{\mu_{2}\nu_{2}}...\Theta^{\mu_{n}\nu_{n}}\,\left[\partial_{\nu_{1}}\partial_{\nu_{2}}...\partial_{\nu_{n}}\,\phi_{2}(x)\right]\,,\label{I-5}
\end{eqnarray}

\noindent 
where $\Theta^{\mu\nu}$ is the antisymmetric real constant
matrix characterizing the noncommutativity of the underlying
space-time \cite{footnote2}.

The gauge fixing is implemented by adding to the action $S_{V}$ the covariant term

\begin{equation}
S_{gf}\,=\,-\frac{a}{2}\int d^{8}z\,\tr\, V\left\{
D^{2},\overline{D}^{2}\right\} V\,,\label{I-10}
\end{equation}

\noindent 
where $a$ is a real constant labeling the gauge. The
corresponding Faddeev-Popov determinant can be cast as

\begin{equation}
\Delta^{-1}\left[V\right]\,=\,\int\mathcal{D}c\,\mathcal{D}c^{\prime}\,\mathcal{D}\overline{c}\,\mathcal{D}\overline{c}^{\prime}\,
e^{iS_{gh}[c,c^{\prime},{\bar c},{\bar c}^{\prime}]}\,.\label{I-12}
\end{equation}

\noindent 
The ghost fields $c,\overline{c}=c^{\dagger},c^{\prime},\overline{c}^{\prime}={c^{\prime}}^{\dagger}$ also take values in the Lie algebra, namely, $c(z) = c_{a}(z) T_{a}$ and so on. The explicit form for the ghost action is found to be

\begin{equation}
S_{gh}\,=\,i\,\left[c+\overline{c}\right]\,L_{\frac{g}{2}V}\,\left[-\left(c+\overline{c}\right)+\left(\coth L_{\frac{g}{2}V}\right)\left[\overline{c}-c\right]\right]\,,
\end{equation}

\noindent 
where

\begin{equation}
L_{A}\left[B\right]\equiv\left[A,B\right]_{*}\,.\label{I-14}
\end{equation}

The ${\cal N}= 2, 4$ supersymmetric theories can be constructed by adding chiral
matter superfields $\Phi^i \left(z\right)\,=\,\Phi^i_{a} \left(z\right)\, T_{a}$.
The action describing a matter superfield interacting with the gauge superfield reads

\begin{equation}
S^{i}_{m}\,=\,\int
d^{8}z\,\tr\,\overline{\Phi}^i*e^{-gV}*\Phi^i*e^{gV}\,.\label{I-19}
\end{equation}

We remind the reader that the self-interaction among the chiral superfields $\Phi^i, i = 1, 2, 3$, entering the maximally extended ${\cal N}= 4$ NCSYM model, may be entirely disregarded as far as the calculation of the one-loop corrections to the n-point vertex functions of the $V$ superfield is concerned. Because of this, the corresponding term has been omitted from the action.

\begin{figure}
\includegraphics{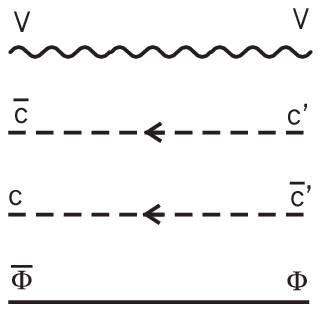}
\caption{\label{propagators}Free propagators.}
\vspace{2cm}
\includegraphics[width=0.70\paperwidth]{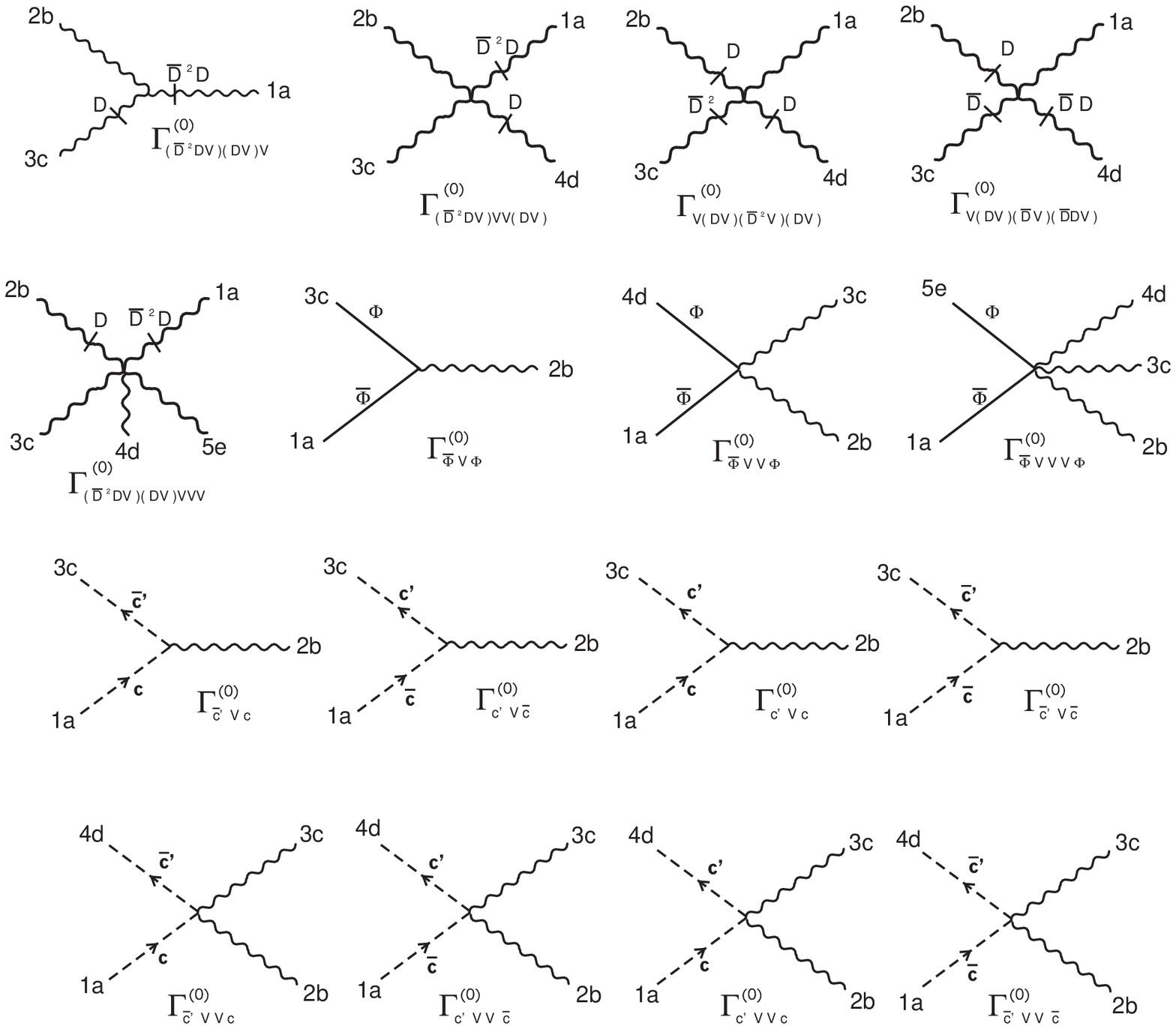}
\caption{\label{vertices}Elementary vertices.}
\end{figure}

From the quadratic part of the action $S_{V}+S_{gf}+S_{gh}+S_{m}$ one obtains the free propagators,

\begin{subequations}
\label{II-1}
\begin{eqnarray}
&&\Delta_{V_a V_b}(z_1 - z_2)\,=\,\delta_{ab}\,\frac{2i}{\Box}\,\left[ 1\,+\,( 1 - a)\,
\frac{1}{\Box}\,\{D_1^2 , {\overline D}_1^2 \} \right]\,\delta^8(z_1 - z_2)\,,\label{mlett:aII-1}\\
&&\Delta_{{\overline c}_a c^{\prime}_b}(z_1 - z_2)\,=\,-\delta_{ab}\,\,\frac{2i}{\Box}\,D_1^2 \,\overline{D}_2^2\,\delta^8(z_1 - z_2) ,\label{mlett:bII-1}\\
&&\Delta_{c_a {\overline c}^{\prime}_b}(z_1 - z_2)\,=\,\delta_{ab}\,\frac{2i}{\Box}\,{\overline D}_1^2 \,{D}_2^2\,\delta^8(z_1 - z_2) \,,\label{mlett:cII-1}\\
&&\Delta_{\Phi^i_a \overline{\Phi}^j_b}(z_1 - z_2)\,=\,-\,\delta^{ij}\,\delta_{ab}\,\,\frac{2i}{\Box}\,{\overline D}_1^2 \,{D}_2^2\,\delta^8(z_1 - z_2)\,,\label{mlett:dII-1}
\end{eqnarray}
\end{subequations}

\noindent 
corresponding to the gauge, ghosts and matter superfields,
respectively. They are depicted in Fig.~\ref{propagators}.
On the other hand, the interacting part of the total action enables
us to find the elementary vertices needed for our calculations.
They are displayed in Fig. \ref{vertices}. In an obvious notation,

\begin{subequations}
\label{II-2}
\begin{equation}
\Gamma_{(\overline{D}^{2}DV_{a})V_{b}(DV_{c})}^{(0)}(k_{1},k_{2},k_{3})\,=\,\frac{ig}{2}\,\mathcal{V}_{3_{abc}}(k_{1},k_{2},k_{3})\,,\label{mlett:aII-2}
\end{equation}
\begin{equation}
\Gamma_{\overline{c}_{a}V_{b}c_{c}^{\prime}}^{(0)}(k_{1},k_{2},k_{3})\,=\,\Gamma_{c_{a}V_{b}\overline{c}_{c}^{\prime}}^{(0)}(k_{1},k_{2},k_{3})\,=\,\frac{ig}{2}\,\mathcal{V}_{3_{abc}}(k_{1},k_{2},k_{3})\,,\label{mlett:bII-2}
\end{equation}
\begin{equation}
\Gamma_{c_{a}V_{b}c_{c}^{\prime}}^{(0)}(k_{1},k_{2},k_{3})\,=\,\Gamma_{\overline{c}_{a}V_{b}\overline{c}_{c}^{\prime}}^{(0)}(k_{1},k_{2},k_{3})\,=\,-\frac{ig}{2}\,\mathcal{V}_{3_{abc}}(k_{1},k_{2},k_{3})\,,\label{mlett:cII-2}
\end{equation}
\begin{equation}
\Gamma_{\overline{\Phi}_{a}V_{b}\Phi_{c}}^{(0)}(k_{1},k_{2},k_{3})\,=\,-ig\,\mathcal{V}_{3_{abc}}(k_{1},k_{2},k_{3})\,,\label{mlett:dII-2}
\end{equation}
\begin{equation}
\Gamma_{(\overline{D}^{2}DV_{a})V_{b}V_{c}(DV_{d})}^{(0)}(k_{1},k_{2},k_{3},k_{4})\,=\,-\frac{ig^{2}}{24}\,\mathcal{V}_{4_{abcd}}^{(1)}(k_{1},k_{2},k_{3},k_{4})\,,\label{mlett:eII-2}
\end{equation}
\begin{equation}
\Gamma_{V_{a}DV_{b}DV_{c}(\overline{D}^{2}V_{d})}^{(0)}(k_{1},k_{2},k_{3},k_{4})\,=\,\frac{g^{2}}{4}\,\mathcal{V}_{4_{abcd}}^{(2)}(k_{1},k_{2},k_{3},k_{4})\,,\label{mlett:fII-2}
\end{equation}
\begin{equation}
\Gamma_{V_{a}DV_{b}\overline{D}DV_{c}(\overline{D}V_{d})}^{(0)}(k_{1},k_{2},k_{3},k_{4})\,=\,-\frac{g^{2}}{4}\,\mathcal{V}_{4_{abcd}}^{(2)}(k_{1},k_{2},k_{3},k_{4})\,,\label{mlett:gII-2}
\end{equation}
\begin{equation}
\Gamma_{c_{a}V_{b}V_{c}\overline{c}_{d}^{\prime}}^{(0)}(k_{1},k_{2},k_{3},k_{4})\,=\,\Gamma_{\overline{c}_{a}V_{b}V_{c}\overline{c}_{d}^{\prime}}^{(0)}(k_{1},k_{2},k_{3},k_{4})\,=\,-\frac{ig^{2}}{12}\,\mathcal{V}_{4_{abcd}}^{(1)}(k_{1},k_{2},k_{3},k_{4})\,,\label{mlett:hII-2}
\end{equation}
\begin{equation}
\Gamma_{\overline{c}_{a}V_{b}V_{c}c_{d}^{\prime}}^{(0)}(k_{1},k_{2},k_{3},k_{4})\,=\,\Gamma_{c_{a}V_{b}V_{c}c_{d}^{\prime}}^{(0)}(k_{1},k_{2},k_{3},k_{4})\,=\,+\frac{ig^{2}}{12}\,\mathcal{V}_{4_{abcd}}^{(1)}(k_{1},k_{2},k_{3},k_{4})\,,\label{mlett:iII-2}
\end{equation}
\begin{equation}
\Gamma_{\overline{\Phi}_{a}V_{b}V_{c}\Phi_{d}}^{(0)}(k_{1},k_{2},k_{3},k_{4})\,=\,\frac{ig^{2}}{2}\,\mathcal{V}_{4_{abcd}}^{(1)}(k_{1},k_{2},k_{3},k_{4})\,,\label{mlett:jII-2}
\end{equation}
\begin{equation}
\Gamma_{\overline{D}^2DV_{a}V_{b}V_{c}V_{d}DV_{e}}^{(0)}(k,p_{1},p_{2},p_{3},-k)\,=\,-\frac{ig^{3}}{24}\,\mathcal{V}_{5_{abcde}}(k,p_{1},p_{2},p_{3},-k)\,,\label{mlett:kII-2}
\end{equation}
\begin{equation}
\Gamma_{\overline{\Phi}_{a}V_{b}V_{c}V_{d}\Phi_{e}}^{(0)}(k,p_{1},p_{2},p_{3},-k)\,=\,-\frac{ig^{3}}{6}\,\mathcal{V}_{5_{abcde}}(k,p_{1},p_{2},p_{3},-k)\,.\label{mlett:lII-2}
\end{equation}
\end{subequations}

\noindent 
Here,

\begin{subequations}
\label{II-3}
\begin{equation}
\mathcal{V}_{3_{abc}}(k_{1},k_{2},k_{3})\,\equiv\,
e^{-ik_{2}\wedge k_{3}}\, A_{abc}-e^{ik_{2}\wedge k_{3}}\,
A_{acb}\,,\label{mlett:aII-3}
\end{equation}
\begin{eqnarray}
\mathcal{V}_{4_{abcd}}^{(1)}(k_{1},k_{2},k_{3},k_{4})\, & \equiv & \, e^{-i\left(k_{1}\wedge k_{2}+k_{3}\wedge k_{4}\right)}A_{abcd}-2e^{-i\left(k_{1}\wedge k_{2}+k_{4}\wedge k_{3}\right)}A_{abdc}\nonumber \\
 &  & +e^{-i\left(k_{1}\wedge k_{4}+k_{2}\wedge
 k_{3}\right)}A_{adbc}\,,\label{mlett:bII-3}
\end{eqnarray}
\begin{equation}
\mathcal{V}_{4_{abcd}}^{(2)}(k_{1},k_{2},k_{3},k_{4})\,\equiv\,
\sin\left(k_{1}\wedge k_{2}\right)\left[e^{-ik_{3}\wedge
k_{4}}A_{abcd}-e^{ik_{3}\wedge
k_{4}}A_{adcb}\right]\,.\label{mlett:cII-3}
\end{equation}
\end{subequations}

\noindent
As for ${\cal V}_{5_{abcde}}$ we shall only be needing ${\cal V}_5$ with two contracted indices, namely,

\begin{eqnarray}
\mathcal{V}_{5_{abc}}(k,p_{1},p_{2},p_{3},-k)\,
\equiv\,A_{dabcd}\,e^{-i\,p_2 \wedge p_3}\,-\,A_{dabdc}\,e^{-i\,p_2 \wedge p_3}-
\nonumber\\
-3\left[ A_{dabdc}\,e^{-i\,p_2 \wedge p_3}\,e^{2i\,k \wedge p_3} -
A_{dadbc}\,e^{-i\,p_2 \wedge p_3}\,e^{-2i\,k \wedge p_1} \right]\,,
\label{mlett:dII-3}
\end{eqnarray}

\noindent 
where $\mathcal{V}_{5_{abc}} \equiv \mathcal{V}_{5_{dabcd}}$. The momenta are taken positive when entering the vertex
and momentum conservation holds in all vertices. We have also introduced
the definition

\begin{equation}
A_{a_{1}\cdots a_{n}}\,\equiv\,\tr \left( T_{a_{1}}\cdots\,T_{a_{n}} \right)\,.\label{II-19}
\end{equation}

\section{One-loop corrections to the two-point vertex function of the gauge superfield}
\label{sec:level3}

We turn now into computing the one-loop corrections to the
two-point vertex function of the $V$ superfield, to be denoted by
$\Gamma_{VV}^{(1)}$. 

\begin{figure}
\includegraphics[width=0.50\paperwidth]{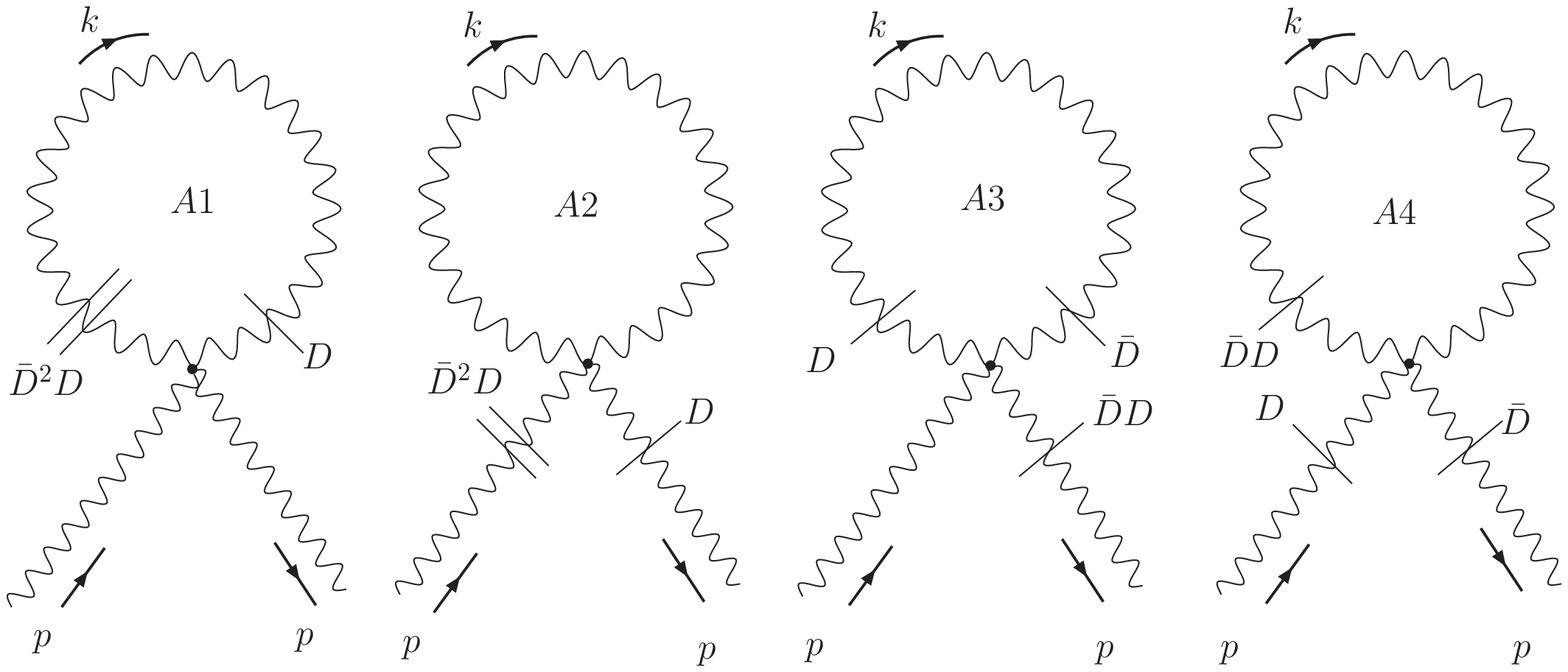}
\caption{\label{tadpoles1}Graphs involving a quartic $V$ vertex.}
\end{figure}

In Fig. \ref{tadpoles1} we draw the contributions containing a quartic
$V$-vertex, omitting those which vanish because of the $D$-algebra alone. According to
the Feynman's rules given in Sec. \ref{sec:level2}, the amplitude 
associated with the graph $A1$ is found to read

\begin{equation}
\Gamma_{A1}\,=\,\left(-\frac{ig^{2}}{24}\right)\int\dk
d^{4}\theta\,\left(F^{A1}\right)_{abcd}\left(-\,\delta^{ad}\frac{2i}{k^{2}}\right)\mathbb{D}_{\theta}^{A1}\,+\,\left(\mbox{sym}\right)\,,\label{III-1}
\end{equation}

\noindent 
where $F^{A1}$ is the phase factor originating from the Moyal
product, $\mathbb{D}_{\theta}$
is the $\theta$-dependent part of the Feynman integrand and $\left(\mbox{sym}\right)$
means symmetrization over external legs which, in this case, implies in
adding the diagram with $p\rightarrow-p$ and $b\leftrightarrow c$.

The calculation of $\mathbb{D}^{A1}_{\theta}$ is elementary and yields

\begin{equation}
\mathbb{D}_{\theta}^{A1}\,=\,-2\,
V^{b}\left(p\right)V^{c}\left(-p\right)\,.\label{III-2}
\end{equation}

\noindent 
The phase factor can be computed by using Eqs. (\ref{mlett:eII-2}) and
(\ref{mlett:bII-3}). Then, one ends up with 

\begin{equation}
\Gamma_{A1}\,=\,\left(\frac{g^{2}}{3}\right)\,\int\dk\dt\frac{1}{k^{2}}\left(F_T\right)_{bc}V^{b}\left(p\right)V^{c}\left(-p\right)\,,\label{III-4}
\end{equation}

\noindent 
where

\begin{equation}
\left(F_T\right)_{bc}\,=\,\left(A_{aabc}\,+\, A_{aacb}\right)\,-\,2cos\left(2k\wedge p\right)A_{abac}\,.\label{III-5}
\end{equation}

\noindent
Notice that $\Gamma_{A1}$ appears to be, by power counting, quadratically UV divergent.

It will prove convenient to introduce the definition

\begin{equation}
Q_{0}\,=\,-\,g^{2}\,\int\dk\dt\frac{1}{k^{2}}\left(F_T\right)_{bc}V^{b}\left(p\right)V^{c}\left(-p\right)\,,\label{III-501}
\end{equation}

\noindent 
which allows us to write

\begin{equation}
\Gamma_{A1}\,=\,-\frac{1}{3}Q_{0}\,.\label{III-502}
\end{equation}

\noindent
According to Eq. (\ref{III-5}), $F_T$ splits into planar (P) and
non-planar (NP) parts. Correspondingly, $Q_{0}=Q_{0}^{P}+Q_{0}^{NP}$.
The planar part $Q_{0}^{P}$ is indeed UV quadratically divergent, whereas $Q_{0}^{NP}$ develops a quadratic infrared pole through the UV/IR mechanism \cite{Minw}.

Next on the line is the graph $A2$. It is not difficult to convince oneself that

\begin{eqnarray}
\mathbb{D}_{\theta}^{A2}\,=\,-\frac{2}{k^{2}}\left(1-a\right)\,\overline{D}^{2}DV^{a}\left(p\right)\,
DV^{d}\left(-p\right)\,,\label{III-7}
\end{eqnarray}

\noindent 
and

\begin{equation}
\left(F^{A2}\right)_{abcd}\delta^{bc}\,=\,\left(A_{ccda}\,+\,A_{ccad}\right)-2e^{-2ik\wedge p}A_{acdc}\,,\label{III-8}
\end{equation}

\noindent 
lead to

\begin{equation}
\Gamma_{A2}\,=\,\left(1-a\right) \left(\frac{g^{2}}{6}\right)\int\dk
d^{4}\theta\,\left[\left(F^{A2}\right)_{abcd}\delta^{bc}\right]\frac{1}{\left(k^{2}\right)^{2}}\,\overline{D}^{2}DV^{a}\left(p\right)\,
DV^{d}\left(-p\right)\,+\,\left(\mbox{sym}\right)\,.\label{III-801}
\end{equation}

\noindent
Hence, $\Gamma_{A2}$ is gauge dependent and contains, at most, logarithmic divergences. To implement the symmetrization with respect to the external legs we first invoke the relation

\begin{equation}
D^{\alpha}\overline{D}^{2}D_{\alpha}V\left(p\right)\,=\,\left(-\not \!
p\,
D\overline{D}+2D^{2}\overline{D}^{2}\right)V\left(p\right)\,=\,\left(+\not \!
p\,\overline{D}D+2\overline{D}^{2}D^{2}\right)V\left(p\right)\,,\label{III-802}
\end{equation}

\noindent 
where $\not\! p\,D {\bar D} = {\not \! p}_{\alpha {\dot \alpha}}\,D^{\alpha} {\bar D}^{{\dot \alpha}}$. Therefore, after realizing that $\left(F^{A2}\right)_{abcd}\delta^{bc}\,+\,\left(\mbox{sym}\right)\,=\,2\left(F_T\right)_{ad}$ one arrives at

\begin{equation}
\Gamma_{A2}\,=\,-\left(\frac{g^{2}}{3}\right)(1-a)\,\int\dk\dt\left(F_T\right)_{ad}\frac{1}{\left(k^{2}\right)^{2}}V^{d}\left(-p\right)\left[-\not \!
p\,
D\overline{D}+2D^{2}\overline{D}^{2}\right]V^{a}\left(p\right)\,.\label{III-11}
\end{equation}

\noindent
However, this expression does not yet appear as being symmetric under the exchange $p\rightarrow-p$ and $a\leftrightarrow d$. In order to explicitly exhibit such symmetry, we start by writing

\begin{eqnarray}
&&\int\dt\, V^{d}\left(-p\right)\left[-\not \! p\, D\overline{D}+2D^{2}\overline{D}^{2}\right]V^{a}\left(p\right) 
\nonumber\\
&& =\frac{1}{2}\int\dt\, V^{d}\left(-p\right)\left[-\not \! p\, D\overline{D}+2D^{2}\overline{D}^{2}\right]V^{a}\left(p\right)\nonumber \\
&& +\,\frac{1}{2}\int\dt\, V^{a}\left(p\right)\left[+\not \! p\,
 D\overline{D}+2D^{2}\overline{D}^{2}\right]V^{d}\left(-p\right)\,,\label{III-12}
\end{eqnarray}

\noindent 
which, after integration by parts in the second term of the right hand side,

\begin{eqnarray}
 &  & \int\dt\, V^{d}\left(-p\right)\left[-\not \! p\, D\overline{D}+2D^{2}\overline{D}^{2}\right]
 V^{a}\left(p\right)
\nonumber \\
  &  & \,\,\,\,=\frac{1}{2}\int \, \dt \,
 V^{d}\left(-p\right)\left[-\not \!  p \left\{ D,\overline{D}\right\}
 +2\left\{ D^{2},\overline{D}^{2}\right\}  \right]V^{a}\left(p\right)\,,\label{III-13}
\end{eqnarray}

\noindent
and since $\left\{ D_{\alpha},{\bar D}_{{\dot \alpha}}\right\} V\left(p\right)={\not \! p}_{\alpha {\dot \alpha}}\, V\left(p\right)$ and ${\not \! p}_{\alpha {\dot \alpha}}\,{\not \! p}^{{\dot \alpha} \alpha}\,=\,2p^{2}$, becomes

\begin{eqnarray}
\int\dt\, V^{d}\left(-p\right)\left[-\not \! p\,
D\overline{D}+2D^{2}\overline{D}^{2}\right]V^{a}\left(p\right) = 
\int\dt\, V^{d}\left(-p\right)\left[-p^{2}+\left\{
D^{2},\overline{D}^{2}\right\}
\right]V^{a}\left(p\right)\,.\label{III-14}
\end{eqnarray}

\noindent
Thus, $\Gamma_{A2}$ can be cast

\begin{equation}
\Gamma_{A2}\,=\,\left(\frac{g^{2}}{3}\right)(1-a)\,\int\dk\dt\left(F_T\right)_{ad}\frac{1}{\left(k^{2}\right)^{2}}V^{d}\left(-p\right)\left[p^{2}-\left\{
D^{2},\overline{D}^{2}\right\}
\right]V^{a}\left(p\right)\,,\label{III-15}
\end{equation}

\noindent 
which is obviously symmetric under the exchange $p\rightarrow-p$ and $a\leftrightarrow d$.

For future purposes, we introduce the definition

\begin{equation}
L_{0}\,=\,g^{2}\,\int\dk\dt\left(F_T\right)_{ad}\frac{1}{\left(k^{2}\right)^{2}}V^{d}\left(-p\right)\left[p^{2}-\left\{
D^{2},\overline{D}^{2}\right\}
\right]V^{a}\left(p\right)\,,\label{III-151}
\end{equation}

\noindent
in terms of which

\begin{equation}
\Gamma_{A2}\,=\,\frac{1}{3}\left(1-a\right)L_{0}\,.\label{III-152}
\end{equation}

As far as the graphs A3 and A4 are concerned, the $D$-algebra yields an integrand \emph{odd} in $k$, in fact, proportional to $\not \! k /k^{4}$. On the other hand, the phase factor is, for both graphs, an \emph{even} function of $k$. Hence, the symmetric $k$-integration leads to

\begin{equation}
\Gamma_{A3}\,=\,\Gamma_{A4}\,=\,0\,.\label{III-16}
\end{equation}

\begin{figure}
\includegraphics[width=0.60\paperwidth]{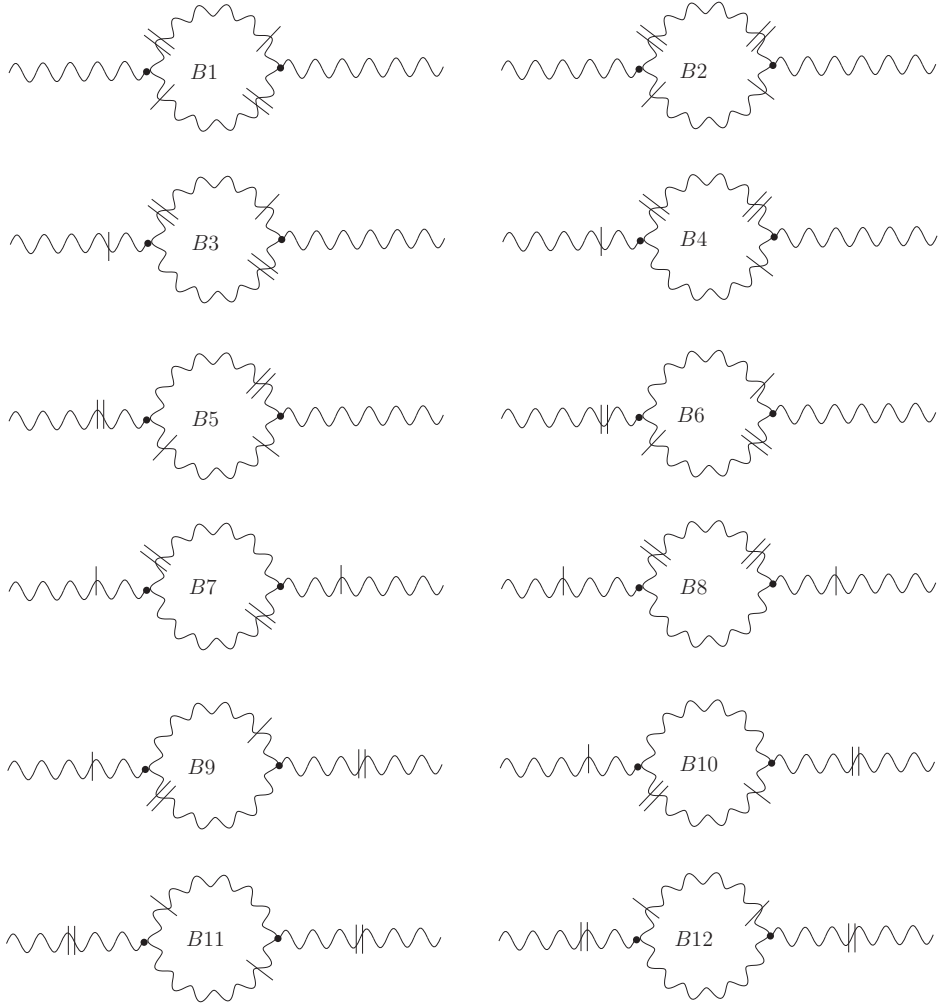}
\caption{\label{vloops}Contributions involving two trilinear $V$ vertices.}
\end{figure}

We next turn into the more complicated task of evaluating the
diagrams involving two trilinear $V$ vertices, depicted in Fig. \ref{vloops}. There
are many graphs to consider, differing among themselves
in the distribution of the $D$-factors in each vertex. For reasons of space, we only present here the details of the calculation of the diagram $B1$. For the remaining ones, we shall merely quote the final results for the corresponding amplitudes.

\begin{figure}
\includegraphics[width=0.40\paperwidth]{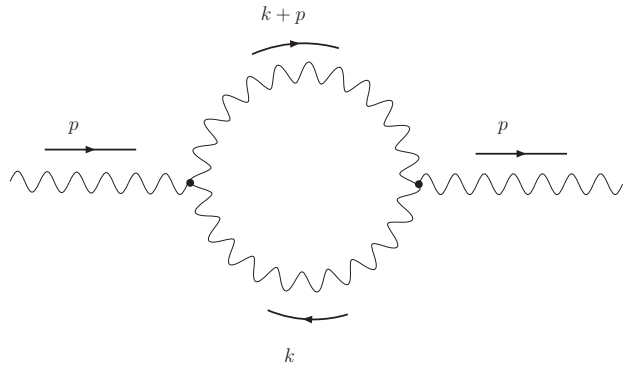}
\caption{\label{vloop-momentum}Momentum flow for the $V$-loop.}
\end{figure}

With the momentum flow indicated in Fig. \ref{vloop-momentum}, the
Feynman rules applied to $B1$ lead to

\begin{equation}
\Gamma_{B1}\,=\,-\,\left(\frac{g^{2}}{2}\right)\int\dk\dt_{1}\dt_{2}\left(F^{B1}\right)_{abcdef}\delta^{af}\delta^{cd}\left[\frac{\left(i\right)^{2}}{k^{2}\left(k+p\right)^{2}}\right]\dth^{B1}\,+\,\left(\mbox{sym}\right)\,.\label{III-17}
\end{equation}

\noindent 
As before, $\left(\mbox{sym}\right)$ means symmetrization over the external legs, i.e.,
adding the expression with $p \rightarrow -p$ and $b \leftrightarrow e$. By feeding the momenta in Fig. \ref{vloop-momentum} into Eq.  (\ref{mlett:aII-3}) one obtains

\begin{equation}
\left(F^{B1}\right)_{abcdef}\delta^{af}\delta^{cd}\,=\,2\,\left(F_{L}\right)_{be}\,,\label{III-18}
\end{equation}

\noindent 
where 

\begin{equation}
\left(F_{L}\right)_{be}\,\equiv\, A_{abc}A_{cea}-cos\left(2k\wedge
p\right)A_{abc}A_{aec}\,.\label{III-181}
\end{equation}

\noindent 
The phase factor $\left(F_{L}\right)_{be}$, common to all diagrams in this topology, is symmetric in both momenta $k$ and $p$ as well as in color indices.

The $D$-algebra for the graph $\Gamma_{B1}$ is identical to that encountered for the corresponding diagram in the $U\left(1\right)$ case \cite{Ferrari1}. Hence, we write at once 

\begin{eqnarray}
\dth^{B1}\, =  \,-2\delta_{12}V_{2}^{e}\left(-p\right)\left[k^{2}\,+\,\not \!
 k\, D\,\overline{D}\,+\,
 D^{2}\overline{D}^{2}\right]V_{1}^{b}\left(p\right)\,.\label{III-19}
\end{eqnarray}

\noindent
By putting all the ingredients together we arrive at

\begin{eqnarray}
\Gamma_{B1}\, & = &
 \,-\,2g^{2}\,\int\dk\dt\left(F_{L}\right)_{be}\left\{ k^{2}\left(\frac{V^{b}\left(p\right)V^{e}\left(-p\right)}{k^{2}\left(k+p\right)^{2}}+\frac{V^{e}\left(-p\right)V^{b}\left(p\right)}{k^{2}\left(k-p\right)^{2}}\right)\,\right.
\nonumber \\
 &  & +\,\not \!
 k\,\left(\frac{D\overline{D}V^{b}\left(p\right)V^{e}\left(-p\right)}{k^{2}\left(k+p\right)^{2}}+\frac{D\overline{D}V^{e}\left(-p\right)V^{b}\left(p\right)}{k^{2}\left(k-p\right)^{2}}\right)
\nonumber \\
 &  &
 \left.+\,\left(\frac{D^{2}\overline{D}^{2}V^{b}\left(p\right)V^{e}\left(-p\right)}{k^{2}\left(k+p\right)^{2}}+\frac{D^{2}\overline{D}^{2}V^{e}\left(-p\right)V^{b}\left(p\right)}{k^{2}\left(k-p\right)^{2}}\right)\right\} \,,\label{III-20}
\end{eqnarray}

\noindent
where the symmetrization has already been performed. 

To isolate different powers of $k$ in the integrand of Eq. (\ref{III-20}), we expand $\left(k\pm p\right)^{-2}$ around $p=0$, i.e.,

\begin{equation}
\frac{1}{\left(k\pm p\right)^{2}}\,=\,\frac{1}{k^{2}}\,\mp2\frac{k\cdot
p}{\left(k^{2}\right)^{2}}\,+\,\frac{4\left(k\cdot
p\right)^{2}-k^{2}p^{2}}{\left(k^{2}\right)^{3}}\,+\,\cdots\,.\label{III-21}
\end{equation}

\noindent 
Then, after some manipulations envolving the $\theta$ integrals one finds

\begin{eqnarray}
\Gamma_{B1}\, & = & \,-\,2g^{2}\,\int\dk\dt\left(F_{L}\right)_{be}V^{e}\left(-p\right)\times\nonumber \\
 &  & \times\left[\frac{2}{k^{2}}+\frac{4\left(k\cdot
 p\right)^{2}-k^{2}p^{2}}{\left(k^{2}\right)^{3}}-4\frac{\left(k\cdot
 p\right)^{2}}{\left(k^{2}\right)^{3}}+\frac{1}{\left(k^{2}\right)^{2}}\left\{ D^{2},\overline{D}^{2}\right\} \right]V^{b}\left(p\right)\,+\, FT\,,\label{III-22}
\end{eqnarray}

\noindent 
where ``FT'' is short for ``finite terms''. To further simplify this expression we first write

\begin{equation}
\Gamma_{B1}\,=\, 2\,Q_{1}\,-\, L_{1}\,+FT\,,\label{III-24}
\end{equation}

\noindent where

\begin{equation}
Q_{1}\,\equiv\,-\,2g^{2}\,\int\dk\dt\frac{1}{k^{2}}\left(F_{L}\right)_{be}V^{b}\left(p\right)V^{e}\left(-p\right)\,\label{III-25}
\end{equation}

\noindent
and

\begin{eqnarray}
L_{1}\,&=&\,2g^{2}\,\int\dk\dt\left(F_{L}\right)_{be}\frac{1}{\left(k^{2}\right)^{2}}V^{e}\left(-p\right)
\nonumber\\
&\times&
\left[\frac{4\left(k\cdot
p\right)^{2}-k^{2}p^{2}}{\left(k^{2}\right)}-4\frac{\left(k\cdot
p\right)^{2}}{\left(k^{2}\right)}+\left\{
D^{2},\overline{D}^{2}\right\}
\right]V^{b}\left(p\right)\,.\label{III-2501}
\end{eqnarray}

\noindent
From observation follows that $Q_{1}$ and $L_{1}$ are, respectively, quadratically and logarithmically divergent by power counting. Furthermore, for the planar part of
$L_{1}$ one can take advantage of

\begin{equation}
\int d^{4}k\,
k_{\mu}k_{\nu}f\left(k^{2}\right)\,=\,\frac{1}{4}g_{\mu\nu}\int
d^{4}k\, k^{2}f\left(k^{2}\right)\label{III-2502}
\end{equation}

\noindent
to write

\begin{equation}
L_{1}^{P}\,\equiv\,-\,2g^{2}\,\int\dk\dt\frac{1}{\left(k^{2}\right)^{2}}\left(F_{L}^{P}\right)_{be}V^{e}\left(-p\right)\left(p^{2}-\left\{
D^{2},\overline{D}^{2}\right\}
\right)V^{b}\left(p\right)\,.\label{III-26}
\end{equation}

\noindent 
The nonplanar part $L_{1}^{NP}$ develops a logarithmic UV/IR infrared
pole which, for being harmless, can be lumped into ``FT''. To summarize, we may cast $\Gamma_{B1}$ in the following final form

\begin{equation}
\Gamma_{B1}\,=\,2Q_{1}-L_{1}^{P}\,+FT\,.\label{III-27}
\end{equation}

\noindent 
The planar part of $Q_{1}$ is quadratically UV divergent, while the nonplanar one
develops a quadratic UV/IR infrared pole. As can be seen from (\ref{III-26}), $L_{1}^{P}$ is logarithmically UV divergent.

For the remaining diagrams in Fig. \ref{vloops} we found

\begin{eqnarray}
&&\Gamma_{B2}\,=\,-2Q_{1},
\hspace{1cm}\Gamma_{B3}\,=\,L_{2}^{P}-2L_{1}^{P} + FT,
\hspace{1cm}\Gamma_{B4}\,=\,0, \nonumber\\
&&\Gamma_{B5}\,=\,2L_{1}^{P} + FT,
\hspace{1cm}\Gamma_{B6}\,=\,-2L_{1}^{P} + FT,
\hspace{1cm}\Gamma_{B7}\,=\,L_{1}^{P} + FT, \nonumber \\
&&\Gamma_{B8}\,=\,-2\left(1-a\right)L_{1}^{P},
\hspace{1cm}\Gamma_{B9}\,=\,-2aL_{1}^{P} + FT,\hspace{1cm}
\Gamma_{B10}\,=\,-2aL_{1}^{P} + FT,
\nonumber\\
&&\Gamma_{B11}\,=\,FT,
\hspace{1cm}\Gamma_{B12}\,=\,FT\,, \label{III-351}
\end{eqnarray}

\noindent
where

\begin{equation}
L_{2}^{P}\,\equiv\,-\,2g^{2}\,\int\dk\dt \frac{p^{2}}{\left(k^{2}\right)^{2}}
\left(F_{L}^{P}\right)_{be}V^{e}\left(-p\right)V^{b}\left(p\right)\,.\label{III-35}
\end{equation}

By adding up Eqs. (\ref{III-502}), (\ref{III-152}), (\ref{III-16}), (\ref{III-27}), (\ref{III-351}), one arrives at

\begin{equation}
\label{sum1}
\Gamma_{A}\,+\,\Gamma_{B}\,=\,-\frac{1}{3}Q_{0} \,+\, \frac{1}{3}(1-a)L_0
\,-\,2(2+a)L_1^P \,+\, L_2^P \,+\,FT\,.
\end{equation}

\begin{figure}
\includegraphics[width=0.35\paperwidth]{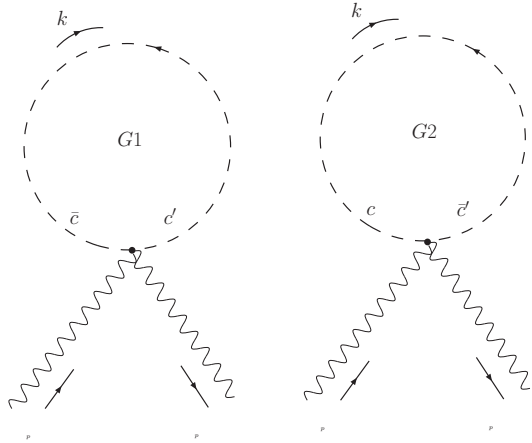}
\caption{\label{gtadpole}Ghost tadpole contributions.}
\end{figure}

Ghost loop contributions to $\Gamma_{VV}^{\left(1\right)}$ involving a quartic vertex arise
from the graphs depicted in Fig.  \ref{gtadpole}. A straightforward application of the Feynman rules yields

\begin{equation}
\Gamma_{G1}\,=\,\left(-1\right)\left(\frac{ig^{2}}{12}\right)\int\dk\dt\,\left(F^{G1}\right)_{abcd}\left(\delta^{ad}\frac{2i}{k^{2}}\right)\dth^{G1}\,+\,\left(\mbox{sym}\right)\,,\label{III-57}
\end{equation}

\noindent 
and

\begin{equation}
\Gamma_{G2}\,=\,\left(-1\right)\left(-\frac{ig^{2}}{12}\right)\int\dk\dt\,\left(F^{G2}\right)_{abcd}\left(-\delta^{ad}\frac{2i}{k^{2}}\right)\dth^{G2}\,+\,\left(\mbox{sym}\right)\,.\label{III-58}
\end{equation}

\noindent 
In view of $F^{G1}=F^{G2}$ and $\dth^{G1}=\dth^{G2}$ one concludes that $\Gamma_{G1}=\Gamma_{G2}$ and, hence,

\begin{equation}
\Gamma_{G1}+\Gamma_{G2}\,=\,-\frac{2}{3}Q_{0}\,.\label{III-6101}
\end{equation}

\begin{figure}
\includegraphics[width=0.65\paperwidth]{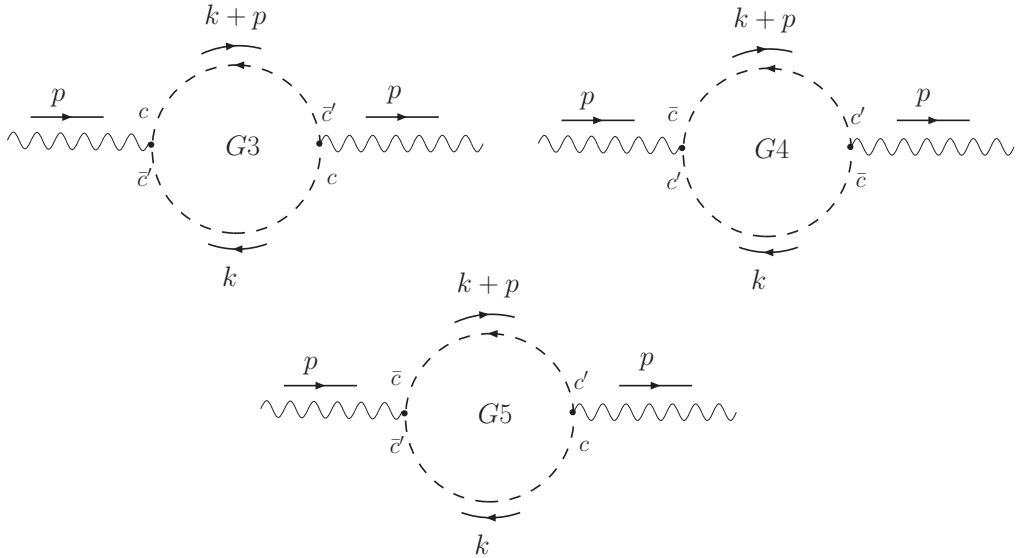}
\caption{\label{gloops}Diagrams involving a ghost loop.}
\end{figure}

The diagrams containing a ghost loop with two trilinear vertices are indicated in Fig. \ref{gloops}. Both $\Gamma_{G3}$ and $\Gamma_{G4}$ are very similar to $\Gamma_{B1}$. Indeed,  $\Gamma_{G3}\,=\,\Gamma_{G4}$ and $\Gamma_{G3}+\Gamma_{G4}=\Gamma_{B1}$. Therefore, according to Eq. (\ref{III-27}),

\begin{equation}
\Gamma_{G3}+\Gamma_{G4}\,=\,2Q_{1}-L_{1}^{P}\,+FT\,.\label{III-65}
\end{equation}

It remains to compute the graph $G5$. Since

\begin{equation}
\dth^{G5}\, = \,
D^{2}\overline{D}^{2}\delta_{12}D^{2}\overline{D}^{2}\delta_{12}V^{b}\left(p\right)V^{e}\left(-p\right)\, = \,\delta_{12}\overline{D}^{2}D^{2}V^{b}\left(p\right)V^{e}\left(-p\right)\,,\label{III-66}
\end{equation}

\noindent 
it can at most be logarithmically divergent. The topological weight for this diagram is $2$. Then,

\begin{equation}
\label{III-681}
\Gamma_{G5}\,=\, L_{3}^{P}\,+FT\,,
\end{equation}

\noindent 
where

\begin{equation}
L_{3}^{P}\,=\,2g^{2}\,\int\dk\dt_{1}\frac{1}{\left(k^{2}\right)}\left(F_{L}^{P}\right)_{be}V^{e}\left(-p\right)\left\{
\overline{D}^{2},D^{2}\right\} V^{b}\left(p\right)\,.\label{III-69}
\end{equation}

The total contribution of ghost loops to $\Gamma_{VV}^{(1)}$
is obtained from Eqs. (\ref{III-6101}), (\ref{III-65}), (\ref{III-681}), and amounts to

\begin{equation}
\label{sum2}
\Gamma_G \, = \, -\frac{2}{3}Q_1 \,-\, L_1^P \,+\, L_3^P \,+\,FT\,.
\end{equation}

\noindent
One should notice that Eqs. (\ref{III-26}), (\ref{III-35}), and (\ref{III-69}) imply that 

\begin{equation}
L_{1}^{P}\,=\, L_{2}^{P}\,+\, L_{3}^{P}\,.\label{III-691}
\end{equation}

\begin{figure}
\includegraphics[width=0.15\paperwidth]{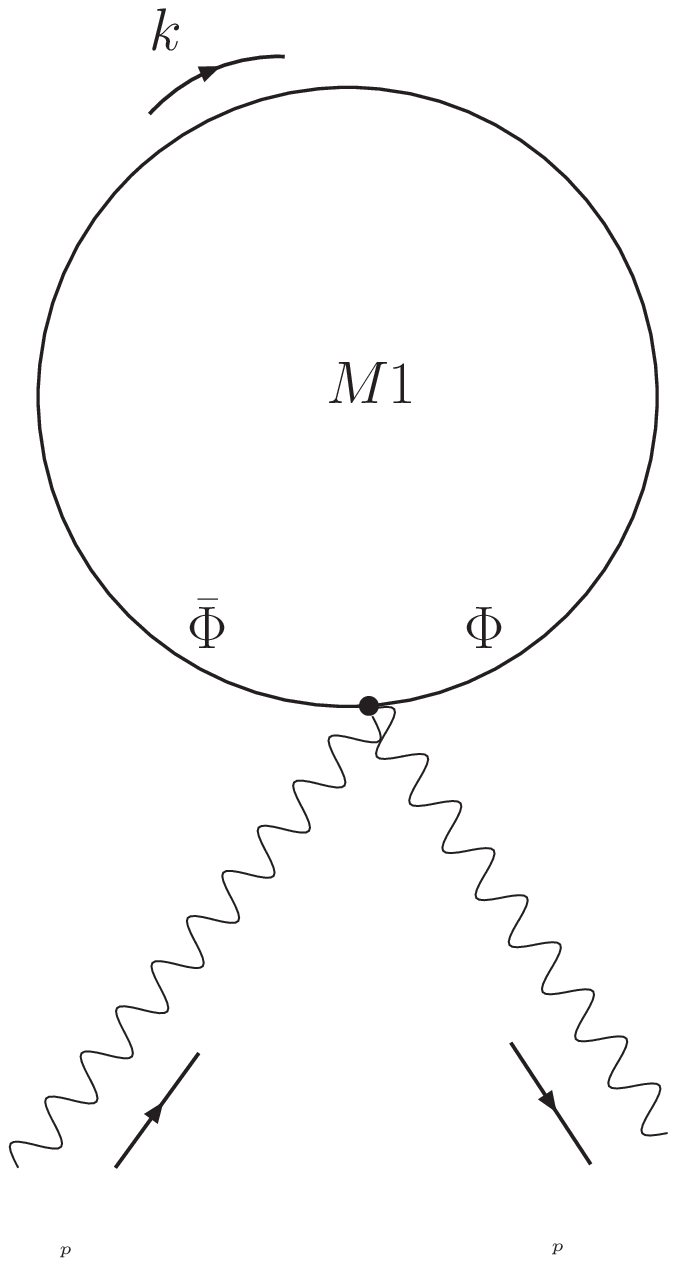}
\includegraphics[width=0.40\paperwidth]{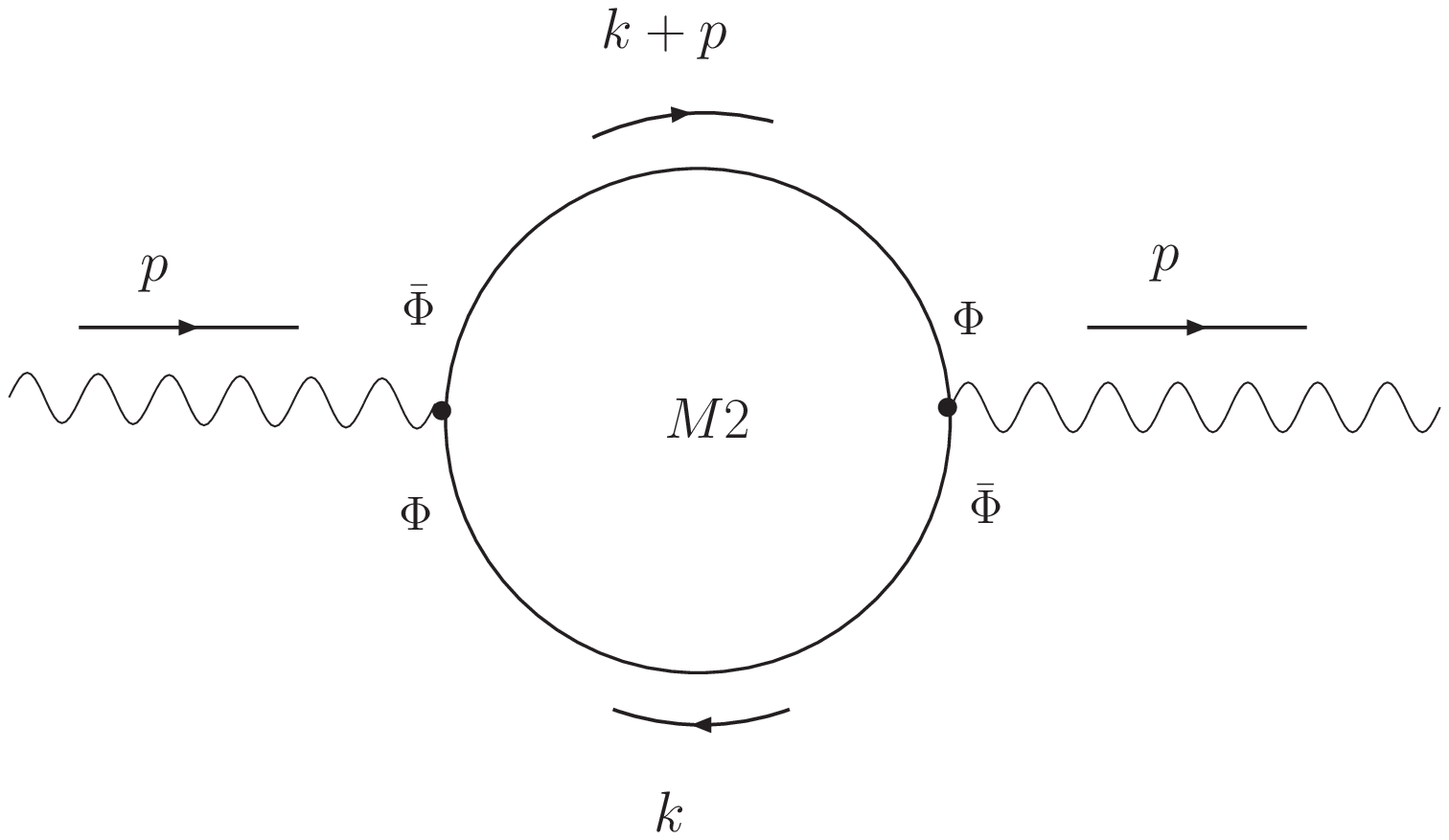}
\caption{\label{mattergraphs}Matter contributions.}
\end{figure}

We address now the calculation of the matter contributions to the
two-point vertex function of the $V$ superfield. The relevant diagrams
for each matter superfield are depicted in Fig. \ref{mattergraphs}. 
Up to numerical factors, their evaluation is just as for the
corresponding ghost graphs, since $c,c^{\prime}$
($\overline{c},\overline{c}^{\prime}$) and
$\Phi^i$ ($\overline{\Phi}^i$) are 
both chiral (antichiral) superfields. Thus, for diagram $M1$ one finds

\begin{equation}
\label{III-72}
\Gamma_{M1}\,=\,2Q_{0}\,.
\end{equation}

\noindent
As for $M2$, one has $\Gamma_{M2}\,=\,-4\Gamma_{G3}\,=\,-2\Gamma_{B1}$, i.e.,

\begin{equation}
\Gamma_{M2}\,=\,-4Q_{1}+2L_{1}^{P}\,+FT\,\label{III-74}
\end{equation}

\noindent
and, therefore,

\begin{equation}
\label{III-741}
\Gamma_{M1} + \Gamma_{M2} \, = \, 2\left(Q_{0} \,-\, 2Q_1 \,+\,L_1^P \right) \,+\, FT \,
\end{equation}

\noindent
is the one-loop correction to $\Gamma_{VV}$ contributed by each chiral matter superfield. In particular, for the maximally extended (${\cal N}=4$) NCSYM theory,

\begin{equation}
\label{sum3}
\Gamma_M \, = \, 3\times(\Gamma_{M1} + \Gamma_{M2}) \,.
\end{equation}

Now we are able to discuss the structure of divergences of
$\Gamma_{VV}^{(1)}$. Let us first focus on its planar part, which
contains all the UV divergences.
Quadratic UV divergences are washed out by dimensional
regularization, while linear ones are killed by symmetric integration.
All that is left are logarithmic UV divergences and UV finite terms. 
For ${\cal N}=1, 2$, these divergences must be renormalized. 
As for ${\cal N}=4$, it turns out that

\begin{equation}
\label{III-75}
\left[ \Gamma_{VV}^{(1)} \right]_{\textrm{logarithmic
UV}} \,=\, 2(1\,-\,a)\,L_1^P \,+\, \frac{1}{3} (1-a) L_0 \,,
\end{equation}

\noindent
as seen from Eqs. (\ref{sum1}), (\ref{sum2}), (\ref{sum3}), and (\ref{III-691}). 
As in the commutative case \cite{Storey1,Storey2}, the maximally extended supersymmetric theory turns out to be free of ultraviolet divergences in the Feynman gauge ($a=1$).

We concentrate next into studying the nonplanar part of
$\Gamma_{VV}^{(1)}$, which due to the noncommutativity is UV finite
but develops infrared poles through the UV/IR mechanism. 
As for the $U\left(1\right)$ model \cite{Ferrari1}, the phase factor originated from the
noncommutativity is always an even function of $k$ and, hence, there can be no linear
UV/IR infrared divergences. Then, the only harmful UV/IR
poles are the quadratic ones, contained in $Q_{1}$ and $Q_{0}$. 
For the pure gauge sector one finds (see Eqs. (\ref{sum1}) and (\ref{sum2}))

\begin{equation}
\label{III-76}
\left[\Gamma_{A}^{NP}+\Gamma_{B}^{NP}+\Gamma_{G}^{NP}\right]_{\textrm{quadratic
UV/IR}}\,=\,-Q_{0}^{NP}+2Q_{1}^{NP}\,,
\end{equation}

\noindent while, for each chiral matter superfield (see Eq.  (\ref{III-741})),

\begin{equation}
\label{III-77}
\left[\Gamma_{M1}^{NP}+\Gamma_{M2}^{NP}\right]_{\textrm{quadratic
UV/IR}}\,=\,2Q_{0}^{NP}-4Q_{1}^{NP}\,.
\end{equation}

\noindent
Therefore, the quadratic UV/IR infrared divergences cancel both for 
the ${\cal N}=1$ as well as for the extended
supersymmetric NCSYM if

\begin{equation}
\label{III-78}
\frac{1}{2}Q_{0}^{NP}\,=\, Q_{1}^{NP}\,,
\end{equation}

\noindent 
which, in view of the definitions in Eqs.(\ref{III-501}) and (\ref{III-25}), demands that

\begin{equation}
\label{III-79}
\int\dk\left(F_T^{NP}\right)_{be}\int\dt
V^{b}\left(p\right)V^{e}\left(-p\right)\,=\,4\,\int\dk\left(F_{L}^{NP}\right)_{be}\int\dt
V^{b}\left(p\right)V^{e}\left(-p\right)\,.
\end{equation}

\noindent 
According to Eqs. (\ref{III-5}) and (\ref{III-181}), a \emph{sufficient} condition 
for Eq.  (\ref{III-79}) to hold is

\begin{equation}
\label{III-80}
\tr \left(T_{a}T_{b}T_{a}T_{e}\right)\,=\,2\,\tr \left(T_{a}T_{b}T_{c}\right)\,\tr\left(T_{a}T_{e}T_{c}\right)\,.
\end{equation}

\noindent
For the fundamental representation of $U(N)$, the set of generators is complete and, therefore \cite{Penati},

\begin{equation}
\label{III-81}
(T_a)_{ij}(T_a)_{kl}\,=\,\frac{1}{2}\,\delta_{il}\delta_{jk}\,,
\end{equation}

\noindent 
which guarantees that Eq. (\ref{III-80}) is, in fact, verified. 

\section{Leading UV/IR infrared divergences in the one-loop
corrections to the three-point vertex function}
\label{sec:level4}

In connection with higher-point vertex functions, it is natural to expect that the cancellation of the nonintegrable UV/IR infrared singularities will require further conditions involving the traces of the group generators, like in Eq. (\ref{III-80}). The natural question is whether these conditions will be verified by the generators in the fundamental representation of the gauge group $U(N)$. A throughout investigation to provide a full answer to this question is clearly impracticable but we may, nevertheless, start to clarify the situation by looking at the one-loop corrections to the three-point vertex function of
the gauge superfield $V$, hereafter to be denoted by $\Gamma_{VVV}^{(1)}$.   
For reasons of simplicity, we shall restrict here to study the leading
(quadratic by power counting) divergences.

\begin{figure}
\includegraphics[width=0.40\paperwidth]{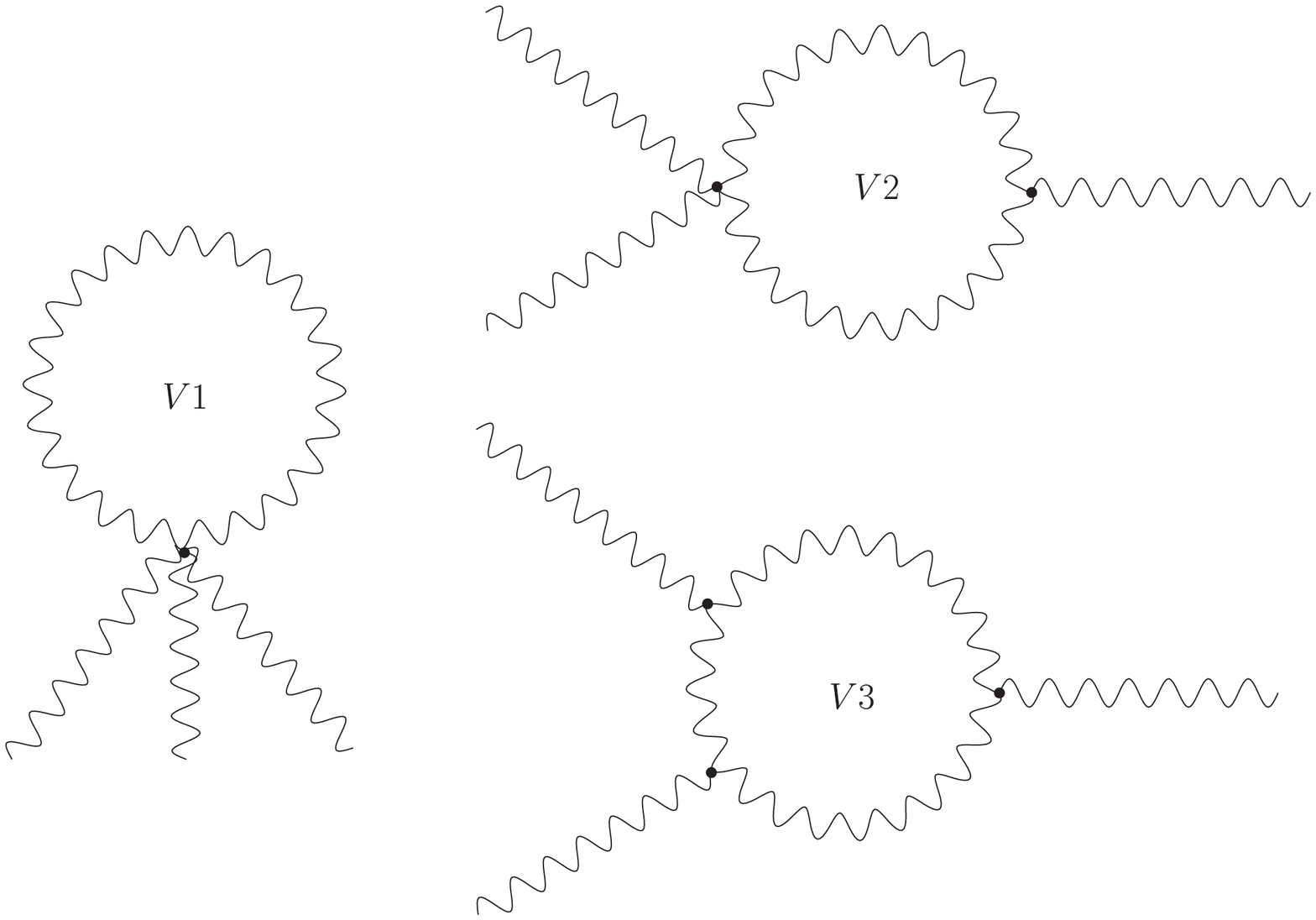}
\caption{\label{vloops3}Vector loop contributions to $\Gamma_{VVV}^{(1)}$.}
\end{figure}

The diagrams contributing to $\Gamma_{VVV}^{(1)}$ and involving a vector loop  
are generically depicted in Fig.~\ref{vloops3}. We shall first address the 
supergraph $V1$ involving the quintic vertex of the gauge
superfield. The corresponding amplitude is found to read

\begin{equation}
\label{IV-1}
\Gamma_{V1} \,=\, \left( - \frac{ig^3}{24} \right) \int \dk \dth \,
\left(F_{V1}\right)_{abcde}\left(-\delta^{ae}\frac{2i}{k^{2}}\right)\mathbb{D}_{\theta}^{G}\,+\,\left(\mbox{sym}\right)\,.
\end{equation}

\noindent
The phase factor $F_{V1}$ can be
calculated from Eq. (\ref{mlett:dII-3}). Since the objects of interest are
the leading UV/IR infrared singularities, we single out the nonplanar part of $F_{V1}$ which is proportional to

\begin{equation}
e^{-i\,p_2 \wedge p_3}\, \left[ A_{dabdc}\,e^{2i\,k \wedge p_3} -
A_{dadbc}\,e^{-2i\,k \wedge p_1} \right]\,.
\end{equation}

\noindent
With the help of 

\begin{equation}
\label{IV-2}
\int\dk\frac{e^{2ik\wedge p}}{k^2}=\frac{1}{4\pi^2 p\circ p}\,
\end{equation}

\noindent 
and after total symmetrization with respect to the external
momenta and color indices, one concludes that $\Gamma_{V1}$ does not contain quadratic UV/IR
infrared divergences.

\begin{figure}
\includegraphics[width=0.40\paperwidth]{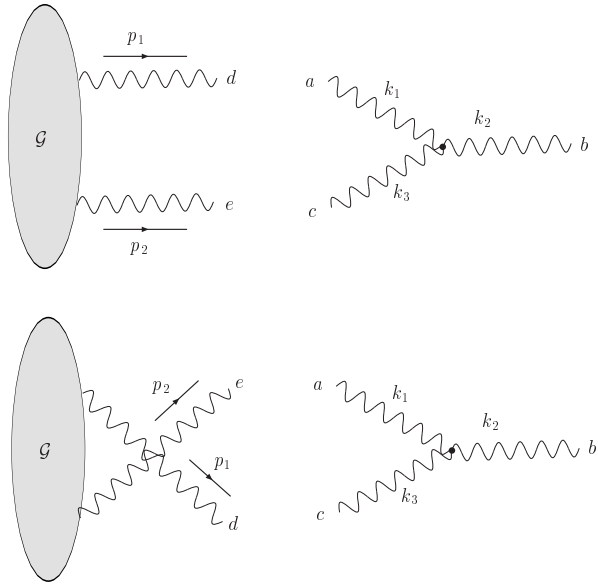}
\caption{\label{trick}Antisymmetric property of the triple vertex $\bar{D}^2DV^a V^b DV^c$. }
\end{figure}

Leading UV/IR infrared divergences arising from the remaining diagrams in 
Fig. \ref{vloops3} also vanish as consequence of a simple
property involving the triple vertex $\bar{D}^2DV^a V^b DV^c$:
the exchange of two legs contracted with the field derivative factors contained in the just mentioned vertex, implies in an overall change of sign in the corresponding amplitude. To understand why this happens, let us consider some (sub)supergraph ${\cal G}$ with two legs $V^d(p_1)$ and $V^e(p_2)$ to be contracted with the $\bar{D}^2DV^a V^b DV^c$ vertex (see Fig. \ref{trick}). The amplitude associated with ${\cal G}$ will schematically be given by

\begin{equation}
\label{IV-3}
(\ldots )_{de}V^d(p_1)V^e(p_2).
\end{equation}

\noindent
We observe that only the terms involving derivates in the triple $V$ vertex are to be contracted with $V^d(p_1)$ and $V^e(p_2)$. Indeed, if doing otherwise we would not be looking at a diagram containing leading divergences. As indicated in Fig. \ref{trick}, there are two ways to perform such contraction:
 
\noindent 
(i) $V^d$ is contracted with $V^a$ and $V^e$ with $V^c$. The
resulting amplitude reads

\begin{equation}
\label{IV-4}
[\ldots ]_{de}\delta^{da}\delta^{ec}[e^{ip_1\wedge
p_2}A_{abc}-e^{-ip_1\wedge p_2}A_{acb}]\,=\,
[\ldots ]_{de}[e^{ip_1\wedge
p_2}A_{dbe}-e^{-ip_1\wedge p_2}A_{deb}]\,.
\end{equation}

\noindent (ii) 
$V^d$ is contracted with $V^c$ and $V^e$ with
$V^a$. In this case the amplitude turns out to be

\begin{equation}
\label{IV-5}
[\ldots ]_{de}\delta^{dc}\delta^{ea}[e^{ip_2\wedge
p_1}A_{abc}-e^{-ip_2\wedge p_1}A_{acb}]\,=\,
-\,[\ldots ]_{de}[e^{ip_1\wedge
p_2}A_{dbe}-e^{-ip_1\wedge p_2}A_{deb}]\,.
\end{equation}

\noindent
Clearly, the sign difference between Eqs. (\ref{IV-4}) and (\ref{IV-5}) is at the root of the mechanism of cancellation for the leading divergent contributions arising from diagrams $V2$ and $V3$ in Fig. \ref{vloops3}.

\begin{figure}
\includegraphics[width=0.30\paperwidth]{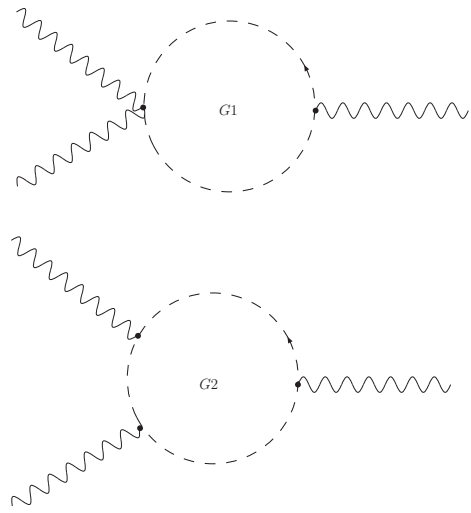}
\caption{\label{gloops3}Ghost loops contributions to $\Gamma_{VVV}^{(1)}$.}
\end{figure}

As for the ghost loop contributions, depicted in Fig. \ref{gloops3},
the cancellation of the leading UV/IR infrared singularities is a
direct consequence of the Feynman rules given in Sec. \ref{sec:level2}.

\begin{figure}
\includegraphics[width=0.40\paperwidth]{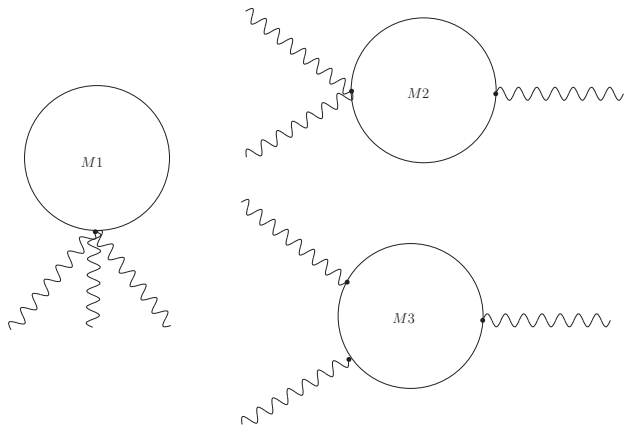}
\caption{\label{mloops3}Matter loops contributions to $\Gamma_{VVV}^{(1)}$.}
\end{figure}

We finally turn into considering the matter loop contributions to
$\Gamma_{VVV}^{(1)}$, shown in Fig. \ref{mloops3}.
The amplitude associated with $M1$ is proportional to the
one corresponding to diagram $V1$ in Fig. \ref{vloops3} and, hence, its nonplanar
part vanishes. 

The phase factor corresponding to the supergraph $M2$, involving one triple and
one quartic matter vertices, is given by

\begin{eqnarray}
\label{IV-6}
\left(F_{M2}\right)_{eabdecd}&=& e^{-i(2k\wedge p_3+p_1\wedge p_2)}A_{eabd}A_{ecd}-e^{-i(-2k\wedge
p_3+p_1\wedge p_2)}A_{dabe}A_{dce}\nonumber\\
&-& 2e^{-i(-2k\wedge p_1-p_1\wedge p_2)}A_{eadb}A_{ecd}+2e^{-i(2k\wedge
p_2-p_1\wedge p_2)}A_{eadb}A_{dce}\,.
\end{eqnarray}

\noindent 
The sum of the first two terms turns out to be an odd
function of the integration momentum $k$ and, therefore, does not contribute to the leading divergences. 
Hence, the nonplanar piece of the amplitude containing the leading UV/IR infrared divergences  is proportional to

\begin{eqnarray}
\label{IV-7}
\int \dk &&\frac{1}{k^2}\left( e^{-i(2k\wedge p_1-p_1\wedge p_2)}A_{eadb}A_{ecd}-
e^{-i(2k\wedge p_2-p_1\wedge p_2)}A_{eadb}A_{dce} \right)
\,+\,(\mbox{sym})\,=\nonumber\\
&& = \,\frac{1}{4\pi^2}\,e^{i\,p_1\wedge p_2}
\left(\frac{A_{eadb}A_{ecd}}{p_1\circ p_1}-\frac{A_{eadb}A_{dce}}{p_2\circ p_2}
\right)\,+\,(\mbox{sym})\,,
\end{eqnarray}

\noindent 
which can be cast

\begin{eqnarray}
\label{IV-8}
\frac{i}{2\pi^2}\sin(p_1\wedge p_2)\,\Big[&&
\frac{1}{p_1\circ p_1}(A_{eadb}A_{ecd}-A_{eadc}A_{ebd})\,+\,
\frac{1}{p_2\circ p_2}(A_{ebdc}A_{ead}-A_{ebda}A_{ecd})
\nonumber\\&&+\,\frac{1}{p_3\circ p_3}(A_{ecda}A_{ebd}-A_{ecdb}A_{ead})\,\Big]\,.
\end{eqnarray}

\noindent 
It is easy to see that this last expression vanishes if and only if

\begin{equation}
\label{IV-9}
\tr(T_dT_aT_eT_b)\tr(T_dT_cT_e)\,=\,\tr(T_dT_aT_eT_c)\tr(T_dT_bT_e)\,.
\end{equation}

\noindent
Again, Eq. (\ref{III-81}) suffices to secure that (\ref{IV-9}) holds. 

Finally, the supergraph $M3$, involving three matter vertices, does not yield
leading UV/IR infrared singularities due to a mechanism similar to the
one described in connection with diagram $V2$ in Fig. \ref{vloops3}.

We then conclude that the restriction to the fundamental group representation protects the quantum corrections to the vertex functions of the $V$ superfield from the appearance of nonintegrable UV/IR infrared divergences. We emphasize that the same condition is required for the $U(N)$ to become an operational gauge group at the classical level \cite{Chaichian}.

\section{Summary}
\label{sec:level5}

In this work we studied the structure of the divergences of
the two- and three-point vertex functions of the $V$ gauge superfield in the
NCSYM theory in four spacetime
dimensions with the aim of establishing the consistency of the model.

In the planar sector, we found logarithmic UV divergences which cancel
in the ${\cal N} = 4$ extended supersymmetric theory in the Feynman
gauge. We confirm the belief
that ${\cal N}=4$ NCSYM theory is finite \cite{JJ}.  

As for the nonplanar sector, the restriction of the group generators 
to be in the fundamental representation turned out to be essential to guarantee the 
cancellation of the leading UV/IR infrared divergences in
the two- and three-point vertex functions of $V$. Therefore, one may hope that 
the UV/IR mechanism will not jeopardize the
perturbative expansion at higher-loop orders. 

We close this paper by pointing out that the results in this work are in agreement with the understanding of noncommutative supersymmetric field theories as the low energy limit of open superstrings in the presence of an external magnetic field \cite{SW}. We first recall that the matrices of the gauge groups associated with the open type I superstring (SST I) are restricted to the fundamental representations of $U(N)$, $SO(N)$ and $USp(N)$ \cite{footnote4}. The low energy limit of SST I is ${\cal N} = 1$ SYM, in ten space-time dimensions, and reduces to ${\cal N} = 4$ SYM when six dimensions are compactified. As known, these field theories are insensitive to the dimension of the gauge group matrices. On the other hand, when a background magnetic field is turned on the low energy limit of the SST I is a noncommutative supersymmetric gauge field theory which only exists in the fundamental representation of the gauge group matrices. In fact, the above restriction turns out to be essential for the quantum corrections to be free of the harmful UV/IR infrared singularities.

\vspace{1cm}

{\bf Acknowledgements.} This work was partially supported by Funda\c
c\~ao de Amparo \`a Pesquisa do Estado de S\~ao Paulo (FAPESP) and
Conselho Nacional de Desenvolvimento Cient\'\i fico e Tecnol\'ogico
(CNPq). H. O. G. and V. O. R. also acknowledge support from PRONEX
under contract CNPq 66.2002/1998-99. A. Yu. Petrov has been supported
by FAPESP, project No. 00/12671-7.

\newpage

\begin{thebibliography}{99}
\bibitem{footnote1} We think that it is not necessary to quote a long list of references concerned with the general aspects of noncommutative quantum field theories. The interested reader may, for instance, consult the list of references in Ref. \cite{Ferrari1}.
\bibitem{footnote3} The noncommutative $SO(N)$ and $USp(N)$ algebras have also already been constructed \cite{Matsubara,Bonora1,Bars}. Nevertheless, this procedure is considerably more involved.
\bibitem{footnote2} To evade unitarity and causality problems we restrict ourselves to the case of space-space noncommutativity ($\Theta_{0\,i}=0$).
\end{thebibliography}
\end{document}